\documentclass[aps,prd,reprint,amsmath,amssymb,superscriptaddress,nofootinbib,longbibliography,floatfix]{revtex4-2}

\usepackage{graphicx}
\usepackage{xcolor}
\usepackage{slashed}
\usepackage[colorlinks=true,linkcolor=blue,citecolor=blue,urlcolor=blue]{hyperref}
\usepackage{bbm}

\newcommand{\ie}{{i.e.}}

\newcommand{\dd}{\mathrm{d}}
\newcommand{\trD}{\mathrm{tr}_{\rm D}}

\newcommand{\Sp}{\mathrm{Sp}}

\begin{document}

\title{Towards Quark-Gluon Correlations in Pion from the Dyson-Schwinger Equations}

\author{Zanbin Xing}
\affiliation{School of Physics, Nankai University, Tianjin 300071, China}

\author{Jiayin Kang}
\affiliation{School of Physics, Nankai University, Tianjin 300071, China}

\author{Lei Chang}
\email[Corresponding author: ]{leichang@nankai.edu.cn}
\affiliation{School of Physics, Nankai University, Tianjin 300071, China}

\date{November 25, 2026}

\begin{abstract}
Twist-3 distributions encode quark-gluon correlations, and the pion's chiral-odd $e(x)$ is the simplest of them. QCD fixes its lowest moments exactly, yet the only rainbow-ladder (RL) calculation of $e(x)$ omitted the Wilson line and violated the equation-of-motion (EOM) sum rule for the first moment. We show that the discrepancy is not a failure of RL but of the omitted link: at first order in the coupling, the link contributes three classes of planar diagrams that cancel the EOM violation class by class, so the sum rule is restored exactly in RL. The resulting zeroth and first moments are complete RL results. The genuine twist-3 component changes sign near $x=0.45$, and the Wilson line more than doubles its magnitude in the negative region. This identifies the Wilson line as the missing ingredient in the RL description of $e(x)$ and sets the stage for a full resummation of the $gA$ expansion.
\end{abstract}

\maketitle

\section{Introduction}
\label{sec:intro}

Parton distribution functions of leading twist are single-particle densities: they count the partons that carry a given light-front momentum fraction, $x$. Higher-twist distributions are not. They express correlations between quark and gluon fields, and so reveal aspects of the binding dynamics that leading-twist distributions cannot~\cite{Jaffe:1991kp,Jaffe:1991ra,Efremov:2002qh}. The chiral-odd, unpolarised twist-3 distribution, $e(x)$, is the simplest of them. Its Mellin moments are forward matrix elements of the local operators $\bar\psi(n\cdot iD)^{j}\psi$, in which the gluon field enters through the covariant derivative. QCD fixes two of these moments: the zeroth moment is the $\sigma$ term divided by the current-quark mass, and, owing to the equation of motion (EOM), the first is $m/M$, where $M$ is the hadron mass~\cite{Efremov:2002qh}. These constraints provide a sharp test of any framework that aims to describe quark-gluon correlations. The EOM relation requires the complete covariant derivative, equivalently the Wilson line in the bilocal operator that defines $e(x)$.

The distribution has a long history. It was introduced in 1991, along with transversity, as one of the chiral-odd distributions accessible in Drell--Yan processes~\cite{Jaffe:1991kp,Jaffe:1991ra}. Its scale dependence was subsequently established~\cite{Koike:1996bs,Belitsky:1997zw}, and $e(x)$ was shown to contain a $\delta(x)$ singularity, so that its moment sum rules receive a contribution invisible to measurements at $x>0$~\cite{Burkardt:2001iy}. The nonperturbative origin of that singularity and the shape of $e(x)$ were then elucidated in the chiral quark-soliton model~\cite{Wakamatsu:2003uu,Schweitzer:2003uy}. Empirically, constraints on $e(x)$ are sparse and, for the pion, absent. For the nucleon, beam-spin asymmetries in dihadron SIDIS from CLAS provide a clean access and have yielded a first extraction, albeit with large uncertainties~\cite{Courtoy:2022kca}. Lattice QCD has not yet constrained the genuine twist-3 part for either hadron~\cite{Bhattacharya:2020jfj}, and measurements at the Electron-Ion Collider are expected to improve the nucleon constraints substantially~\cite{AbdulKhalek:2021gbh}.

Continuum Schwinger function methods have long been used to calculate the pion's leading-twist distributions---see, e.g., Refs.~\cite{Hecht:2000xa,Chang:2014lva,Bednar:2018mtf,Ding:2019qlr,Cloet:2018ukf,Ding:2020syk}. Recently, these methods were applied to $e(x)$ within the rainbow-ladder (RL) truncation~\cite{Shi2602}. That analysis obtained both the twist-2 $f(x)$, and twist-3 $e(x)$ from a single integral equation for the forward $\pi\pi$ scattering amplitude with two quark legs (constructed as in Ref.~\cite{Xing:2022mvk}), and found that the genuine twist-3 part dominates $xe(x)$ and changes sign. However, the Wilson line is omitted in Ref.~\cite{Shi2602}. Consequently, Ref.~\cite{Shi2602} obtained $\langle x\rangle_e=0.113$, whereas the QCD EOM relation requires $m/m_\pi=0.039$, and attributed the mismatch to a failure of RL to preserve the action of the QCD EOM on the correlator.

Here we show that the discrepancy is instead attributable to the omitted Wilson line. We incorporate it order by order in $g$; at first order, a gluon emitted from the Wilson line is absorbed within the pion in one of three planar diagrams, and we find that these contributions cancel, class by class, the EOM violation of the link-free calculation, so that the EOM sum rule holds in RL. Because the $j$th moment contains at most $j$ powers of the gauge field, only the zeroth and first moments are complete at the order computed here, and the $x$-dependence of $xe^{\rm tw3}$ inferred from moments $j\ge1$ is therefore indicative only. A full resummation of the $g_A$ expansion is in principle feasible within RL but lies beyond the scope of the present work, which should therefore be read as a first step towards a complete RL description of $e(x)$.

This paper is organised as follows. Section~\ref{sec:def} gives the definition and moment expansion of $e(x)$; Sec.~\ref{sec:rl} presents the RL truncation and the sum-rule argument; Sec.~\ref{sec:res} reports the results; and Sec.~\ref{sec:summary} places them in the wider context.

\section{Definition, moments, and $g$ expansion}
\label{sec:def}

In this section we work in Minkowski space, where the definitions and the exact relations are conventionally written~\cite{Jaffe:1991kp,Jaffe:1991ra,Efremov:2002qh}.

The scalar distribution is defined as follows~\cite{Jaffe:1991kp,Jaffe:1991ra}:
\begin{equation}
e(x)=\frac{n\cdot P}{M}\int\frac{d\lambda}{4\pi}\,e^{i\lambda x\,n\cdot P}\,
\langle P|\bar\psi(0)\,\mathcal{W}[0,\lambda n]\,\psi(\lambda n)|P\rangle\,,
\label{EqDef}
\end{equation}
where $|P\rangle$ denotes the pion state with momentum $P$, $n$ is a light-like vector, and $x=n\cdot k/n\cdot P$ is the light-front momentum fraction carried by the struck quark with momentum $k$. $\mathcal{W}[0,\lambda n]{}=\mathcal P\exp\big(-ig\int_0^1\dd s\,\lambda n\cdot A(s\lambda n)\big)$ is the straight Wilson line that maintains the gauge invariance of the bilocal operator, with $D_\mu=\partial_\mu-igA_\mu$. We consider the $u$ quark in the $\pi^+$ and omit the flavour label. 

The Mellin moments of Eq.\eqref{EqDef}, $\langle x^j\rangle_e\equiv\int_{-1}^{1}dx\,x^j e(x)$, can be translated into forward matrix elements of local operators~\cite{Mezrag:2022pqk}:
\begin{equation}
\langle x^j\rangle_e=\frac{\langle P|\bar\psi\,(n\cdot iD)^j\psi|P\rangle}{2M\,(n\cdot P)^j}\,.
\label{EqMom}
\end{equation}
The covariant derivative carries the gauge field, and it is through $D$ that the gluon content of the Wilson line enters each moment. 

The zeroth moment is the scalar density, and by virtue of the Feynman--Hellmann theorem,
\begin{equation}
\langle x^0\rangle_e=\frac{\langle P|\bar\psi\psi|P\rangle}{2M}
=\frac{\sigma_\pi}{2m}
=\frac{1}{4M}\frac{\partial M^2}{\partial m}\,,
\label{EqSigma}
\end{equation}
where $m=m_u=m_d$ is the current-quark mass and $\sigma_\pi=m\,\partial M/\partial m$. The QCD EOM, $(i\slashed D-m)\psi=0$, and its Dirac conjugate, together with $\tfrac12\{\slashed n,i\slashed D\}=n\cdot iD$, gives $\langle P|\bar\psi\,n\cdot iD\,\psi|P\rangle=m\,\langle P|\bar\psi\,\slashed n\,\psi|P\rangle$, and hence~\cite{Efremov:2002qh}
\begin{equation}
\langle x\rangle_e=\frac{m}{M}\,\langle x^0\rangle_f\,,
\label{EqEOM}
\end{equation}
where $\langle x^0\rangle_f=1$ is the quark number obtained from the twist-2 distribution $f(x)$.

The same analysis yields the decomposition~\cite{Efremov:2002qh,Belitsky:1997zw}
\begin{equation}
e(x)=\frac{\langle P|\bar\psi\psi|P\rangle}{2M}\,\delta(x)+e^{\rm mass}(x)+e^{\rm tw3}(x)\,,
\label{EqDecomp}
\end{equation}
in which $x\,e^{\rm mass}(x)=(m/M)\,f(x)$ and $e^{\rm tw3}(x)$ is the genuine twist-3 (quark-gluon) component, whose two lowest moments vanish. The $\delta(x)$ term in Eq.~\eqref{EqDecomp} makes no contribution to hard processes. Therefore, following Ref.~\cite{Shi2602}, we concentrate on
\begin{equation}
\langle x^j\rangle_{xe^{\rm tw3}}=\langle x^{j+1}\rangle_e-\frac{m}{M}\,\langle x^j\rangle_f\,,\qquad
\langle x^0\rangle_{xe^{\rm tw3}}=0\,.
\label{EqXeTw3}
\end{equation}
The second identity follows from Eq.~\eqref{EqEOM} and is the sum rule restored in this work. Throughout this work, $f(x)$ is computed without the link; hence $\langle x^j\rangle_f$ denotes the link-free twist-2 moment.

Acting on a quark of momentum $k$, $\psi\propto e^{-ik\cdot z}$, the combination $iD_\mu$ becomes $k_\mu+gA_\mu$, so the $j$th moment is constructed from $[n\cdot(k+gA)]^{j}$. In powers of $g$, one has
\begin{align}
&\big[n\cdot iD\big]^j\to(n\cdot k)^j+\sum_{a+b=j-1}(n\cdot k_1)^a\,g\,n\cdot A\,(n\cdot k_2)^b \nonumber\\
&\quad+\!\!\sum_{a+b+c=j-2}\!\!(n\cdot k_1)^a\,g\,n\cdot A\,(n\cdot k_2)^b\,g\,n\cdot A\,(n\cdot k_3)^c\nonumber\\
&\quad+\ldots+\mathcal{O}(g^j)\,,
\label{EqExpand}
\end{align}
where the last momentum in each term ($k_2$ at first order, $k_3$ at second order) is that of $\psi$, and each preceding momentum has absorbed one additional gluon. Upon division by $(n\cdot P)^{j}$, the weight at first order is given as:
\begin{equation}
W^{(1)}_j=\sum_{a+b=j-1}x_1^a x_2^b\,,
\label{EqWeights}
\end{equation}
with $x_i=n\cdot k_i/n\cdot P$. For $j=0$ the expansion contains no power of $A$, so the zeroth moment receives no link contribution at any order. More generally, $[n\cdot(k+gA)]^{j}$ contains at most $j$ powers of $A$, so the $j$th moment terminates at order $g^{j}$. We write $\langle x^j\rangle_e=\sum_{k=0}^{j}\langle x^j\rangle_e^{(k)}$ with the index $k$ denotes contributions of order $g^k$.

\section{Rainbow-ladder truncation and the EOM sum rule}
\label{sec:rl}
Hereafter we work in Euclidean space where the our calculation is formulated\footnote{ The product of four vectors $a,\, b$ is $a\cdot b=\sum_{i=1}^{4}a_i b_i$ and $\{\gamma_\mu,\gamma_\nu\}=2\delta_{\mu\nu}$ with $\gamma_\mu^\dagger=\gamma_\mu$.}.
The mapping from Minkowski space to Euclidean space can be achieved by require  
$a^E_4=ia_M^0$, $a^E_i=a_M^i$, $\gamma_4^E=\gamma^0_M$ and $i\gamma_i^E=\gamma_M^i$, and hence $(a\cdot b)_M=-(a\cdot b)_E$ and $(\gamma\cdot a)_M=-i(\gamma\cdot a)_E$.

\subsection{Truncation and exact moments}
We work within the rainbow-ladder truncation in Landau gauge, employing the infrared component of the Qin--Chang interaction~\cite{Qin:2011dd,Maris:1999nt}
\begin{equation}
\mathcal{G}(\ell^2)=\frac{8\pi^2 D}{\omega^4}\,e^{-\ell^2/\omega^2}\,,\qquad
T_{\rho\sigma}(\ell)=\delta_{\rho\sigma}-\frac{\ell_\rho\ell_\sigma}{\ell^2}\,.
\label{EqQC}
\end{equation}
Since this interaction is ultraviolet finite, no renormalisation is necessary ($Z_2=Z_4=1$), and $m$ denotes the Lagrangian mass. The dressed-quark propagator, $S(k)=-i\gamma\cdot k\,\sigma_V(k^2)+\sigma_S(k^2)$, is obtained as the solution of
\begin{equation}
\begin{aligned}
S^{-1}(k)&=S_{0}^{-1}(k)+\Sigma(k)\,,\\
\Sigma(k)&=\int_q C_F\mathcal{G}(\ell^2)\,T_{\rho\sigma}(\ell)\,\gamma_\rho S(q)\gamma_\sigma\,,
\end{aligned}
\label{EqGap}
\end{equation}
where $\ell=k-q$, $C_F=4/3$, and $S_0^{-1}(k)=i\gamma\cdot k+m$ is the bare inverse propagator.

The pion Bethe--Salpeter amplitude, $\Gamma(k;P)$, expressed in terms of the quark momentum $k$ (with the antiquark carrying $k-P$), is obtained by solving the meson BSE
\begin{equation}
\Gamma(k;P)=-\int_q C_F\mathcal{G}(\ell^2)\,T_{\rho\sigma}(\ell)\,\gamma_\rho \,\chi(q;P)\,\gamma_\sigma
\label{EqBSE}
\end{equation}
and is normalised through canonical condition \cite{Maris:2000sk}. The Bethe--Salpeter wave functions are $\chi(k)=S(k)\Gamma(k;P)S(k-P)$.

The amputated forward $\pi\pi$ amplitude, $\mathcal{F}(k,P)$ sums the $t$-channel ladder in addition to the Born term $\mathcal{F}_0(k,P)=\Gamma(k;P)S(k-P)\bar\Gamma(k;-P)$, and can be obtained by solving the following equation:
\begin{equation}
\mathcal{F}(k,P)=\mathcal{F}_0(k,P)-\int_q C_F\mathcal{G}(\ell^2)\,T_{\rho\sigma}(\ell)\,\gamma_\rho\Phi(q,P)\gamma_\sigma\,,
\label{EqF}
\end{equation}
with $\Phi(q,P)=S(q)\mathcal{F}(q,P)S(q)$.

The amplitude $\Phi$ can then be used to express the link-free moments, \ie,Eq.~\eqref{EqMom} with $D\to\partial$:
\begin{gather}
\langle x^j\rangle_e^{(0)}=-\frac{N_c}{M}\trD\int_k\Big(\frac{n\cdot k}{n\cdot P}\Big)^j\Phi(k,P),\nonumber\\
\langle x^j\rangle_f=-\frac{N_c}{n\cdot P}\trD\int_k\Big(\frac{n\cdot k}{n\cdot P}\Big)^j i\,n\cdot\gamma\,\Phi(k,P)\,,
\label{EqMom0}
\end{gather}
where $\langle x^j\rangle_f$ is the link-free twist-2 moment. $\trD$ stands only for Dirac trace.

\begin{figure*}[htbp!]
\centering
\includegraphics[width=0.7\textwidth]{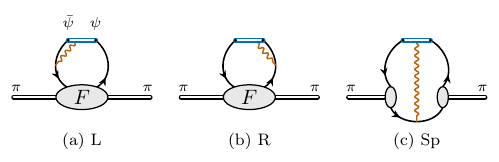}
\caption{\label{FigDiagrams}
Planar Wilson-line diagrams in RL truncation, first order in $g$.
Legend:
blue double line -- straight Wilson line between $\bar\psi$ (left end) and $\psi$ (right end);
solid lines with arrows -- dressed-quark propagators;
wavy lines -- gluon exchange;
hollow double lines -- pions;
large ellipse -- forward $\pi\pi$ amplitude, $F$;
small ellipses -- pion Bethe--Salpeter amplitudes.
}
\end{figure*}

\subsection{Wilson-line attachment in rainbow ladder}

The moments of the full bilocal operator are obtained by replacing $n\cdot k$ with $n\cdot(k+gA)$ in Eq.~\eqref{EqMom0}, the expansion being Eq.~\eqref{EqExpand}. For $j=1$ the expansion terminates at first order, so the first-order Wilson line is a defining part of the moment, not a correction. 
A field insertion on a quark line, with the gluon absorbed where RL permits\footnote{The gluon may attach only where it does not cross another gluon.}, yields
\begin{equation}
\bar\psi\,M_\rho\,gA_\rho\,\psi\ \longmapsto\ -i\int C_F\mathcal{G}(\ell^2)\,T_{\rho\sigma}(\ell)\;M_\rho\otimes\gamma_\sigma\,.
\label{EqInsert}
\end{equation}
Apparently, the gap equation is consistent with Eq.~\eqref{EqInsert}. While for the Wilson-line attachment, at the order considered here, a single gluon is emitted from the link, and terminates with three possibilities: the left leg before the first rung, on the right leg after the last rung, or, in the Born term, on the spectator line between the two pion amplitudes. These three possibilities are labeled as  L, R and Sp, respectively, and are depicted in Fig.~\ref{FigDiagrams}. Every other attachment gives crossed diagrams, which RL omits for the same reason that it omits crossed ladders.

With $M_\rho=n_\rho\mathbbm{1}$ and the weight in Eq.~\eqref{EqWeights}, one obtains
\begin{align}
M\langle x^j\rangle_e^{(1)}={}&iN_c\sum_{X}\int_{k_1}\!\int_{k_2}C_F\mathcal{G}(\ell^2)\,T_{\rho\sigma}(\ell)\,\frac{n_\rho}{n\cdot P}\nonumber\\
&\times W^{(1)}_j(x_1,x_2)\,\trD\mathcal{D}^X_\sigma(k_1,k_2)\,,
\label{EqLink1}
\end{align}
with the three classes $X=L,R,\Sp$ defined geometrically above, whose Dirac structures $\mathcal{D}_\sigma^X$ are given below
with $\ell=k_1-k_2$ and
\begin{gather}
\mathcal{D}^L_\sigma=S(k_1)\gamma_\sigma\Phi(k_2)\,,\quad
\mathcal{D}^R_\sigma=\Phi(k_1)\gamma_\sigma S(k_2)\,,\nonumber\\
\mathcal{D}^{\Sp}_\sigma=\chi(k_1)\gamma_\sigma\bar\chi(k_2)\,.
\label{EqD}
\end{gather}Here the pion momentum $P$ is suppressed for convenience.
For $j\ge2$, Eq.~\eqref{EqLink1} only gives the first-order link contribution, a prediction of the truncation at this order; the moment itself is not complete until order $g^j$. In the next, we will show how the link-free moments violates the sum rule and how Eq.~\eqref{EqLink1} cancels the violation part class by class.

\subsection{Restoration of the EOM sum rule}
We start from the following identity derived from Dirac algebra:\begin{equation}
n\cdot k=\frac{1}{2i}\{n\cdot\gamma,S_0^{-1}(k)\}+i\,m\,n\cdot\gamma\,.\end{equation}
Inserting this into the link-free $\langle x\rangle_e^{(0)}$ and employing $S_0^{-1}\Phi=\mathcal{F}S-\Sigma\Phi$ and $\Phi S_0^{-1}=S\mathcal{F}-\Phi\Sigma$, one obtains the identity\begin{gather}
M\langle x\rangle_e^{(0)}=m\,\langle x^0\rangle_f+I\,,\nonumber\\
I= \frac{-N_c\trD\int n\cdot\gamma\,(\mathcal{F}S-\Sigma\Phi)+(S\mathcal{F}-\Phi\Sigma)\,n\cdot\gamma}{2i\,n\cdot P}\,.
\label{EqIdef}
\end{gather}
Thus the link-free moment differs from $m\langle x^0\rangle_f/M$ by a remainder $I$, to be analyzed in the following. 

Substituting Eqs.~\eqref{EqGap}--\eqref{EqF} into Eq.~\eqref{EqIdef}, $I$ becomes a sum of six two-loop integrals: the self-energy together with the first ladder rung on each leg, and the two Born terms in which one pion amplitude is opened by means of its Bethe--Salpeter equation. Since $\{n\cdot\gamma,\gamma_\rho\}=2n_\rho$, the six terms pair into the three classes $X=L,R,\Sp$ of
\begin{align}
I={}&\frac{N_c}{in\cdot P}\sum_{X=L,R,\Sp}\int_{k_1}\!\int_{k_2}C_F\mathcal{G}(\ell^2)\nonumber\\
&\times n_\rho T_{\rho\sigma}(\ell)\,\trD\mathcal{D}^X_\sigma(k_1,k_2)\,,
\label{EqIpair}
\end{align}
For $j=1$, $W^{(1)}_1=1$, thus for each class $X=L,R,\Sp$, comparison of Eqs.~\eqref{EqLink1} and \eqref{EqIpair} yields
\begin{equation}
M\langle x\rangle^{(1)}_{e,X}=-I_X\;\Rightarrow\;
M\big(\langle x\rangle_e^{(0)}+\langle x\rangle_e^{(1)}\big)=m\,\langle x^0\rangle_f\,.
\label{EqSumRule}
\end{equation}
The equality holds separately in each class: the link-free calculation alone violates Eq.~\eqref{EqEOM}, and the first-order Wilson line cancels the remainder $I$ class by class, restoring the sum rule without fine-tuning. Since only the shared kernel, the translational invariance of the loop integrals, the Born inhomogeneous term of Eq.~\eqref{EqF}, and the symmetry of $T_{\rho\sigma}$ are used, 
Eq.~\eqref{EqSumRule} holds in any covariant gauge in which all equations are solved consistently.

\section{Results and discussion}
\label{sec:res}

We use the parameters of Ref.~\cite{Shi2602}, $\omega=0.5\,$GeV, $(D\omega)^{1/3}=0.82\,$GeV and $m=5.5\,$MeV, which give $m_\pi=0.1409\,$GeV and $f_\pi=0.0892\,$GeV on the finest grid. Selected numerical details are given in Appendix~\ref{sec:num}.

\subsection{Moments}
\label{SecResBench}
Table~\ref{TabMoments} lists the moments that are computed herein.
We obtain $\langle x^0\rangle_e=6.433$, i.e.\ $\sigma_\pi=2m\langle x^0\rangle_e=0.0708\,$GeV, against $6.45$ and $0.071\,$GeV in Ref.~\cite{Shi2602}. This is close to $m_\pi/2$, as expected for RL near the chiral limit~\cite{Maris:1999nt,Flambaum:2005kc}, and the link-free moments agree with those of Ref.~\cite{Shi2602} to within 0.5\% (Table~\ref{TabMoments}). The Feynman-Hellmann relation, $M\langle x^0\rangle_e=\tfrac14\partial m_\pi^2/\partial m$, is numerically confirmed as an exact property of the truncation. The gauge-link-free part of the calculation is therefore under control.

For the first moment, the link-free calculation fails the EOM sum rule: $M\langle x\rangle^{(0)}_e-m=0.0105\,$GeV, roughly twice the current-quark mass, since $\langle x\rangle^{(0)}_e=0.1136$ against $m/m_\pi=0.039$. The first-order link is a cancellation between $2L$ and Sp ($L=R$ by charge conjugation): for $j=1$, $\langle x\rangle^{(1)}_{e,L}=\langle x\rangle^{(1)}_{e,R}=-0.8699$ and $\langle x\rangle^{(1)}_{e,\Sp}=1.6652$, which sum to $-0.0746$. The EOM sum rule is therefore restored class by class by the first-order Wilson line. In addition, the first order link term changes sign between $j=1$ and $j=2$ and is 72--75\% of the link-free moment for $j=2$--$4$.

\begin{table}[t]
\caption{\label{TabMoments}
Moments (dimensionless), including the first-order Wilson line. $\langle x^j\rangle^{(1)}_e$ is the link sum, $2L+\Sp$, and ``total'' denotes $\langle x^j\rangle^{(0)}_e+\langle x^j\rangle^{(1)}_e$.
For $j=1$ the total equals $m/m_\pi$ ($0.03905$). The last row is the link-free $f$.} 
\begin{ruledtabular}
\begin{tabular}{lcccccc}
$j$ & 0 & 1 & 2 & 3 & 4 & 5 \\ \hline
$\langle x^j\rangle^{(0)}_e$ & 6.433 & 0.1136 & 0.3027 & 0.2777 & 0.2308 & 0.1903 \\
$\langle x^j\rangle^{(1)}_e$ & 0 & $-0.07460$ & 0.2179 & 0.2073 & 0.1720 & (0.141) \\
total & 6.433 & 0.03905 & 0.5206 & 0.4850 & 0.4028 & 0.3317 \\
$\langle x^j\rangle_f$ & 1.000 & 0.3717 & 0.2019 & 0.1294 & 0.0909 & (0.068)
\end{tabular}
\end{ruledtabular}
\end{table}

\subsection{Genuine twist-3 component}
\label{SecResTw3}

Table~\ref{TabXeTw3} records the moments of Eq.~\eqref{EqXeTw3} in two scenarios: without the link (1a) and with the first-order link (1b). In both, $\langle x^n\rangle_f$ is the link-free moment of Table~\ref{TabMoments}, and $m/m_\pi=0.03905$. Only $n=0$ is exact, by the EOM sum rule; the moments $n\ge1$ are indicative, since they involve moments of $e(x)$ beyond the first order link. Without the link the zeroth moment is $0.0746$, which violates the sum-rule; with the first-order link, the restoration of the EoM sum rule is numerically confirmed. For $n\ge1$ the first-order link raises the moments by 75--77\%. Ref.~\cite{Shi2602} estimated $xe^{\rm mass}$ as $0.113\,f(x)$; here the mass term is exactly $(m/m_\pi)f(x)$, and the first moment of $xe^{\rm tw3}$ vanishes because of the Wilson line.

\begin{table}[t]
\caption{\label{TabXeTw3}
Moments of $xe^{\rm tw3}$, Eq.~\eqref{EqXeTw3}: 1a without the Wilson line and 1b with the first-order link. In both, the twist-2 moments are link-free. A parenthesised entry changes by more than $10^{-3}$ between the two finest grids.}
\begin{ruledtabular}
\begin{tabular}{lcc}
$n$ & 1a & 1b \\ \hline
0 & 0.07460 & 0 \\
1 & 0.2882 & 0.5060 \\
2 & 0.2698 & 0.4771 \\
3 & 0.2258 & 0.3978 \\
4 & 0.1867 & 0.3282 \\
5 & 0.1559 & (0.273) \\
6 & 0.1320 & (0.232)
\end{tabular}
\end{ruledtabular}
\end{table}

\begin{figure}[t]
\centering
\includegraphics[width=0.9\columnwidth]{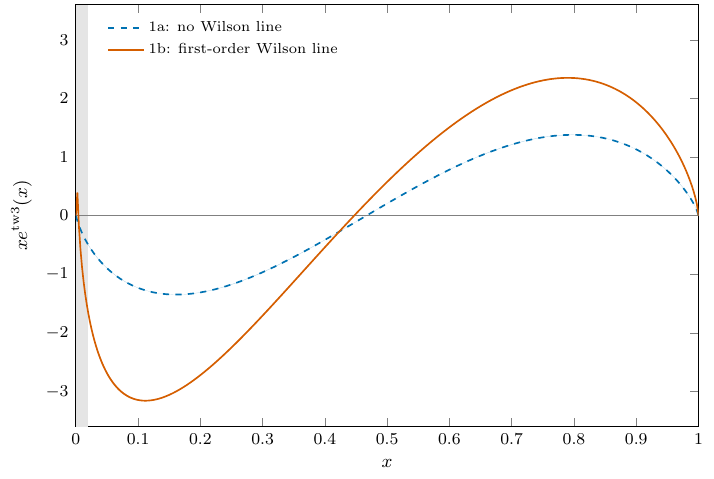}
\caption{\label{FigXe}
Genuine twist-3 component, $xe^{\rm tw3}(x)=xe(x)-(m/m_\pi)f(x)$.
Legend:
dashed blue curve -- without the Wilson line (1a);
solid vermillion curve -- with the first-order link (1b).
The grey strip, $x<0.02$, marks the domain on which the moments do not constrain the curves.}
\end{figure}

Although a finite set of moments does not determine the $x$-dependence. Following Ref.~\cite{Shi2602}, on the support $0\le x\le1$ (no antiquarks are present at the model scale) we fit the complete $xe(x)$, omitting the $\delta(x)$ term, and fit $f(x)$ separately:
\begin{subequations}
\label{EqFit}
\begin{align}
xe(x)&=N(1-x)^b\big[x^a-2x^c(1-x)\big]\,,\label{EqFitXe}\\
f(x)&=(1-x)^\beta\big(c_0+c_1x+c_2x^2\big)\,.\label{EqFitF}
\end{align}
\end{subequations}
The parameters minimise the sum of squared relative deviations from the moments and are listed in Table~\ref{TabFit}. Scenario 1a fits seven moments of $xe^{\rm tw3}$, and scenario 1b fits the five converged moments, $n\le4$. Including the two non-converged moments $n=5,6$ deepens the 1b minimum from $-3.16$ to $-3.74$ but leaves the node and maximum essentially unchanged, and the two curves differ by up to $0.76$ on $x\ge0.02$; the five-moment fit, used for the curves in Fig.~\ref{FigXe}, gives the minimum quoted below.

\begin{table}[t]
\caption{\label{TabFit}
Parameters of Eqs.~\eqref{EqFit}, the number of moments used ($n_{\rm m}$) and the maximal relative deviation from them (dev.). Rows 1a and 1b fit $xe^{\rm tw3}$ directly; the row of Ref.~\cite{Shi2602} is its fit of the complete $xe$. The $f$ row is link-free; it is listed for reference, and the curves in Fig.~\ref{FigXe} do not use it.}
\begin{ruledtabular}
\begin{tabular}{lcccccc}
 & $n_{\rm m}$ & $N$ & $a$ & $b$ & $c$ & dev. \\ \hline
1a & 7 & 8.773 & 0.4997 & 0.7713 & 0.5843 & $2.5\times10^{-3}$ \\
1b & 5 & 11.89 & 0.0894 & 0.6896 & 0.2145 & $4.5\times10^{-4}$ \\
Ref.~\cite{Shi2602} & 8 & 8.493 & 0.4496 & 0.7528 & 0.5512 & $2.6\times10^{-3}$ \\ \hline
 & $n_{\rm m}$ & $\beta$ & $c_0$ & $c_1$ & $c_2$ & dev. \\ \hline
$f$ & 7 & 1.142 & 2.007 & $-2.358$ & 5.767 & $9.7\times10^{-3}$
\end{tabular}
\end{ruledtabular}
\end{table}

The resulting $xe^{\rm tw3}$ is shown in Fig.~\ref{FigXe}. Both scenarios exhibit the shape reported in Ref.~\cite{Shi2602}: $xe^{\rm tw3}$ is negative at small $x$ and positive at large $x$, with a single node. Without the link (1a) the minimum is $-1.35$ at $x=0.16$, the maximum is $1.38$ at $x=0.80$, and the node lies at $x=0.467$. With the first-order link (1b) the five-moment fit gives a minimum of $-3.16$ at $x=0.11$, i.e.\ more than a factor of two deeper, the maximum is $2.35$ at $x=0.79$, and the node lies at $x=0.447$. The link therefore barely shifts the node and the maximum, but it deepens the negative lobe by a factor of 2.3 and raises the positive lobe by a factor of 1.7. The bands propagated from the successive-grid differences of the moments are at most $0.004$ (1a) and $0.025$ (1b) wide on $x\ge0.02$.

On $x\lesssim0.02$ the fits are not constrained by the moments: for $a<c$, Eq.~\eqref{EqFitXe} is dominated by $Nx^a$ as $x\to0$, which for 1b ($a\approx0.09$) produces a narrow positive peak on $x\lesssim0.01$. This peak is an artefact of the form, not a result.

\section{Summary and perspective}
\label{sec:summary}

Using the rainbow-ladder (RL) truncation with the infrared interaction of Ref.~\cite{Shi2602}, we have incorporated the Wilson line of the defining operator into the calculation of the pion's chiral-odd twist-3 distribution, $e(x)$. At first order in the coupling, the link contributes three classes of planar diagrams (Fig.~\ref{FigDiagrams}), which cancel the equation-of-motion violation of the link-free calculation class by class. The QCD sum rule $\langle x\rangle_e=m/m_\pi$ is therefore valid in RL. The zeroth and first moments of $e(x)$ are complete RL results; moments beyond them require higher orders of the $g_A$ expansion and are indicative only.

With the link included, $xe^{\rm tw3}$ retains the shape of Ref.~\cite{Shi2602}---negative at small $x$, a node near $x=0.45$, positive at large $x$---but its zeroth moment now vanishes, as QCD requires, and its negative lobe is more than twice as deep (Fig.~\ref{FigXe}). The twist-2 distribution $f$ that enters $xe^{\rm mass}$ is defined without the Wilson line, Eq.~\eqref{EqMom0}.

A central point of this work is that, in the RL approximation, our method guarantees the relations dictated by the QCD equations of motion. As a concrete check, the spin--orbital correlations reported in Ref.~\cite{Lorce:2025ayr} are reproduced exactly within the present framework. Accordingly, the contribution of this work is to identify the omitted Wilson line as the origin of the EOM mismatch in the earlier RL study and to show that RL is consistent with the QCD EOM. Its boundary is equally clear: because only the zeroth and first moments of $e(x)$ are complete, the moments $j\ge2$ and the large-$x$ shape are indicative; because the interaction is the infrared part of the Qin--Chang model without renormalisation, the results pertain to a model scale; and because planar diagrams saturates the RL truncation, the crossed attachments of link gluons suppressed by $1/N_c^2$ are omitted. Within these limits, the link adds 72--75\% to the $j=2$ moment of $e(x)$, so the quark-gluon correlations that distinguish twist-3 from twist-2 distributions are likely substantial, and the forward amplitude and link classes are independent of the Dirac structure at the operator, so link terms for other structures follow with new traces only.

The natural next step is the completion of the higher moments, which would require a treatment of the straight Wilson line that is not pursued order by order, for instance an integral equation for the link-dressed quark line or a calculation on the light front. Beyond the present calculation, T-odd distributions such as the Boer--Mulders function have been evaluated with a first-order gauge link in the Nambu--Jona-Lasinio model~\cite{Noguera1508} and with a symmetry-preserving contact interaction~\cite{Cheng2409}; within the present framework, they would instead call for the imaginary part of the eikonal propagator. Viewed in this way, the present calculation is a first step towards a complete RL description of \(e(x)\), and it fixes the framework within which the remaining orders can be computed.

\begin{acknowledgments}
Work supported by the National Natural Science Foundation of China (Grant No.~12135007).
\end{acknowledgments}

\appendix{}
\section{Numerical details}
\label{sec:num}

\subsection{Numerical stability}
\label{AppGrid}
The Bethe--Salpeter equation, Eq.~\eqref{EqF} and all moments are evaluated in the pion rest frame. The struck-quark momentum is written $k=p+P/2$, so that the spectator carries $k-P=p-P/2$, and the grid is laid in the real relative momentum $p$: $N_s$ Gauss-Legendre points span $p^2$$\,\in[10^{-4},40]\,$GeV$^2$ under a logarithmic mapping, $N_z$ Chebyshev points the angle between $p$ and $P$, and $N_y$ Gauss-Legendre points the second angle, with grids ranging from $(32,12,12)$ to $(160,48,48)$. The three finest grids are $(N_s,N_z,N_y)=(96,32,32)$, $(128,40,40)$ and $(160,48,48)$. 

Class Sp is evaluated with both loop momenta on the Bethe--Salpeter nodes, so that its discretisation matches that of the Bethe--Salpeter equation term by term and the identity $M\langle x\rangle^{(1)}_{e,\Sp}=-I_{\Sp}$ holds to within rounding. For classes L and R, the same placement leaves a sum-rule residual that barely decreases under grid refinement and stalls near 1\% of $m$. This is because the Gaussian in Eq.~\eqref{EqQC} must be resolved in $q$ when the inner integration variable is the quark momentum. We therefore integrate L and R over the gluon momentum $\ell=k_1-k_2$ on a separate grid, $(48,16,16)$ with $\ell^2\in[10^{-4},40]\,$GeV$^2$. For the three finest grids, the gluon-momentum grid is $(64,24,24)$.

A moment is regarded as numerically stable when its relative change between successive grids is below $10^{-3}$. The stability holds up to $j=7$ for $\langle x^j\rangle^{(0)}_e$, up to $j=5$ for the total moment, and up to $j=4$ for $\langle x^j\rangle_f$. Neither the gluon-momentum grid nor the gap-equation grid limits the precision; the moments $j\le4$ change by $\le2\times10^{-5}$ under further refinement.
On the finest grid, Feynman-Hellmann relation, $M\langle x^0\rangle_e=\tfrac14\partial m_\pi^2/\partial m$, is satisfied to a relative $4\times10^{-6}$ difference; the residual $\Delta=M(\langle x\rangle^{(0)}_e+\langle x\rangle^{(1)}_e)-m$ decreases steadily under grid refinement, reaching $9\times10^{-7}$ of $m$; the unity of quark number is satisfied up to $|\langle x^0\rangle_f-1|<10^{-5}$.

\subsection{Evaluation of the moments}
\label{AppMom}
The moments are evaluated by replacing the dependence on the light-like vector $n$ with a harmonic polynomial in $k$, so that every integral depends on invariants alone. Because $n^2=0$, the ratio $(n\cdot k)^j/(n\cdot P)^j$ is a homogeneous polynomial of degree $j$ in $n$, and rotational invariance about $P$ allows it to be replaced by any harmonic polynomial of degree $j$ that equals unity at $k=P$. We use
\begin{equation}
\frac{(n\cdot k)^j}{(n\cdot P)^j}\mapsto h_j(k)=\Big(\frac{\sqrt{k^2P^2}}{P^2}\Big)^{j}\frac{U_j(z)}{j+1}\,,\quad z=\frac{k\cdot P}{\sqrt{k^2P^2}}\,,
\label{EqHarm}
\end{equation}
where $M^2=-P^2>0$, $U_j$ is the Chebyshev polynomial of the second kind, and the same branch of the square root is taken in both places; since $U_j(-z)=(-1)^jU_j(z)$, $h_j$ is a polynomial in $k\cdot P$ and $k^2P^2$ and does not depend on the branch. Further, $h_j(P)=1$ and $h_1=k\cdot P/P^2$. Since $(n\cdot k)^j$ is harmonic when $n^2=0$, this replacement preserves every integral. The two-loop weight follows from $W^{(1)}_j\,n\cdot V/n\cdot P=\int_0^1\dd s\,V\cdot\partial_y(n\cdot y)^j/(n\cdot P)^j$ at $y=sk_1+(1-s)k_2$, so that
\begin{equation}
\frac{n_\rho}{n\cdot P}W^{(1)}_j\ \mapsto\ \int_0^1\dd s\,\partial_\rho h_j(y)\big|_{y=sk_1+(1-s)k_2}\,,
\label{EqHarmLink}
\end{equation}
a polynomial in $s$ that Gauss-Legendre quadrature integrates exactly. The twist-2 moments use $(n\cdot V)(n\cdot k)^j/(n\cdot P)^{j+1}\mapsto V\cdot\partial h_{j+1}(k)/(j+1)$. Both $h_j$ and $\partial h_j$ are evaluated for all $j$ through the recurrence of $U_j$, which avoids the cancellations of an explicit sum over powers of $k\cdot P$. The price is the factor $h_j\sim(|k|/m_\pi)^j$: the moments are small differences between large numbers, and the discretisation error grows by roughly one power of $|k|/m_\pi$ per unit of $j$, so each grid refinement yields about one additional converged moment.

\bibliographystyle{apsrev4-2}
\bibliography{refs}

\end{document}